# Model Predictive Control of Shallow Drowsiness: Improving Productivity of Office Workers

Takuma Kogo[1], Masanori Tsujikawa[2], Yukihiro Kiuchi[1], Atsushi Nishino[3], and Satoshi Hashimoto[3]

*Abstract*—This paper proposes a methodology of model predictive control for alleviating shallow drowsiness of office workers and thus improving their productivity. The methodology is based on dynamically scheduling setting values for air conditioning and lighting to minimize drowsiness level of office workers on the basis of a prediction model that represents the relation between future drowsiness level and combination of indoor temperature and ambient illuminance. The prediction model can be identified by utilizing state-of-the-art drowsiness estimation method. The proposed methodology was evaluated in regard to a real routine task (performed by six subjects over five workdays), and the evaluation results demonstrate that the proposed methodology improved the processing speed of the task by 8.3% without degrading comfort of the workers.

## I. Introduction

Although automation technologies driven by artificial intelligence and robotics have been replacing some jobs of workers, they does cannot replace all of jobs for some time in the future [1]. Therefore, schemes that support activities involved in production by workers must be continuously devised. Various approaches to support such work-production activities are available; as one approach, providing an appropriate work environment can contribute to workers achieving high productivity.

Previous fundamental studies [2][3][4][5] clarified the relation between work efficiency and indoor temperature (IDT) or ambient illuminance (AMI). The author also assumes the possibility of improving productivity of office workers by utilizing air-conditioning (AC) and lighting (LT)[4]. Especially, the author assumes that the productivity can be improved by alleviating drowsiness during work time. Drowsiness level (DL) can be quantified from eyelid motion of a worker in real-time by estimation methods [6][7][8][9] which have become more practical in recent years.

Previous studies [10][11] reported that a subject was made more awake when the ambient environment had either low IDT or high AMI. It was also reported that a subject who was less drowsy achieved higher performance when doing basic tasks [11][12]. As part of an investigation, the author previously measured DL of workers doing routine tasks with an estimation method proposed in a previous study [6]. Figure 1 shows a histogram of DL of workers for each routine task, and it shows that DL in the range of 1.5 to 2.5 approximately accounts for 40% of the total. This result indicates that workers are not always awake but do their task while shallowly drowsy. Considering these results, the author believe that productivity of office worker can be improved by alleviating this shallow drowsiness.

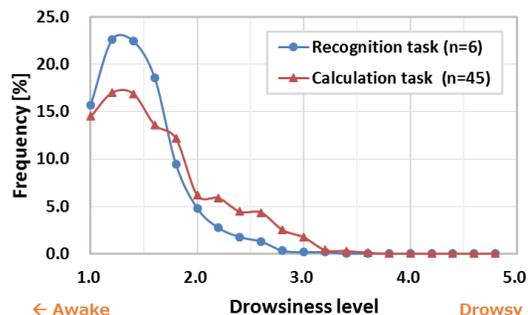

Figure 1. Histogram of worker's drowsiness level for each routine task.

To the best knowledge of the author, a technology for automatically controlling ambient environment with AC and LT on the basis of monitored DL to improve productivity of office has not been studied. Moreover, model predictive control (MPC) [13][14] which can systematically achieve high control performance with multiple control equipment (e.g., AC and LT) and needing less tuning has not been applied. Neither conventional classical feedback nor rule-based control, which are designed with experimental trial-and-error and technical knowhow, can provide these advantage of MPC.

In the present study, the author thus aimed to create a control methodology for minimizing DL of office workers by adjusting AC and LT on the basis of MPC framework. The proposal methodology is dynamically computing schedule of setting values of AC and LT from hour to hour to minimize DL by using prediction models of DL, IDT, and AMI. The prediction model for DL which represents the relation between future DL and combination of IDT and AMI can be identified with past data of them as training data.

The contribution of this paper is twofold: presenting a unprecedented control methodology for shallow drowsiness based on an MPC framework; and evaluating the methodology in terms of objectively measured productivity of office workers in consideration of their comfort (hereinafter, the methodology is called "D-MPC"). A real routine task, as opposed to quasi tasks, was used for the evaluation. The evaluation results show that processing speed measured as objective index is increased. The author believe that these evaluation results show not only basic technical data but also the effectiveness at the point of utility when D-MPC is introduced into a real workplace such as an office.

---

[1] Takuma Kogo and Yukihiro Kiuchi are with Data Science Research Laboratories, NEC Corporation, Japan (e-mail: t-kogo@ah.jp.nec.com).
[2] Masanori Tsujikawa is with Biometrics Research Laboratories, NEC Corporation, Japan.
[3] Atsushi Nishino and Satoshi Hashimoto are with Technology and Innovation Center, Daikin Industries, LTD, Japan.

[4] When becoming more efficient, a worker can obviously do many more tasks or complete them more quickly. Moreover, a worker can also achieve better performance in regard to creative non-routine tasks with the benefit of time and/or mental space additionally obtained [19][20]. It can be expected that making routine tasks more efficient will lead to better performance in regard to whole work because routine and non-routine tasks coexist generally for most workers and a certain amount of routine tasks is inevitably done, although there are various types of business category and assignment.

This paper is structured as follows. In Section II, the proposed control system and its methodology are described. In Section III, the evaluation method is described in detail. In Section IV, the evaluation results are presented and discussed. Finally, in Section V, conclusions derived from these results are drawn.

## II. PROPOSAL METHODOLOGY

### A. System architecture and control policy

Figure 2 shows the architecture of the proposed control system, which executes D-MPC to alleviate shallow drowsiness. The edge device captures and analyzes eyelid motion of a worker via the web camera installed on the head of individual operated monitor in an office environment. From the result of that analysis, it estimates the value of DL [6].The edge device sends the estimated value of DL and values of IDT and AMI measured by sensor placed on the office desk to the server. The server executes calculation of D-MPC, which determines setting values of AC and LT by using received data as current state. The server sends these calculated setting values to the AC and LT. The AC and LT then operate according to the setting values received from the server.

As for the proposed control system, IDT and AMI are adopted as control inputs of D-MPC because available control elements are setting of values of IDT for AC and the same of brightness for LT in common office buildings. Airflow (i.e., volume/direct), radiation heat, color temperature, aroma, and sound were excluded from the control elements because of difficulty to operate them for individual comfort and/or their high system cost.

In D-MPC, schedule of the setting values for AC and LT are calculated at a constant interval by using the latest estimated/measured values of DL, IDT, and AMI. In particular, the schedule can be calculated as not only IDT kept low and/or AMI kept high but also IDT and/or AMI temporally raised to lowered and vice versa by considering DL transition in the future. These changes of IDT and/or AMI have possibility to interact drowsiness as stimulus for awakening. Furthermore, D-MPC runs proactively to prevent workers from getting drowsy even though their DLs are still low. By contrast, the conventional naive controls work at higher DL which is set as a threshold, and it is difficult for them to run proactively.

### B. Details of model predictive control

This subsection describes the detailed calculation method of D-MPC introduced in the previous subsection. The calculation of D-MPC is formulated as a mathematical programming problem defined as a combination of the following (1)-(8), where (1) is an objective function, (2) to (4) are vector expressions regarding the schedule of the setting values for AC and LT, (5) to (7) are prediction models for DL, IDT, and AMI, and (8) is a constraint condition regarding comfort of workers.

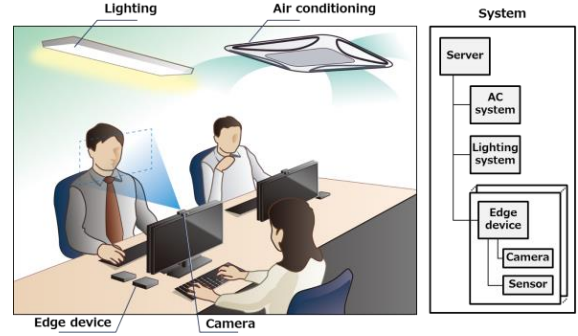

Figure 2. System architecture

$$\boldsymbol{U}^* = \underset{\underline{\boldsymbol{U}} \leq \boldsymbol{U} \leq \overline{\boldsymbol{U}}}{argmin} \frac{1}{\mathcal{W}\mathcal{T}} \sum_{i=1}^{\mathcal{W}} \sum_{t=1}^{\mathcal{T}} D_{i,t}, \quad (1)$$

where $D_{i,t}$ is time-averaged predicted value of DL of worker $i$ at time step $t$-th, where the size of the time step is given as constant $\tau$; $\mathcal{W}$ is number of workers, $\mathcal{T}$ is number of time steps; $\boldsymbol{U}$ is a vector representing the schedule of the setting values for AC and LT; $\overline{\boldsymbol{U}}$ and $\underline{\boldsymbol{U}}$ which are configured in advance as setting parameters are respectively the upper and lower bound of $\boldsymbol{U}$; $\boldsymbol{U}^*$ is $\boldsymbol{U}$ minimizing averaged $D_{i,t}$ for all workers and time steps within the range between the upper and the lower bound, $\overline{\boldsymbol{U}}$ and $\underline{\boldsymbol{U}}$.

$$\boldsymbol{U} = \left[T_1^{(s)}, \ldots, T_{\mathcal{T}}^{(s)}, L_1^{(s)}, \ldots, L_{\mathcal{T}}^{(s)}\right]^{\top}, \quad (2)$$

where $T_t^{(s)}$ and $L_t^{(s)}$ are respectively the setting value of IDT for the AC and AMI for the LT at time step $t$-th.

$$\underline{\boldsymbol{U}} = \left[\underbrace{\underline{T}^{(s)}, \ldots, \underline{T}^{(s)}}_{\mathcal{T}}, \underbrace{\underline{L}^{(s)}, \ldots, \underline{L}^{(s)}}_{\mathcal{T}}\right]^{\top}, \quad (3)$$

where $\underline{T}^{(s)}$ and $\underline{L}^{(s)}$ are respectively the lower bound of the setting value of IDT for the AC and AMI for the LT.

$$\overline{\boldsymbol{U}} = \left[\underbrace{\overline{T}^{(s)}, \ldots, \overline{T}^{(s)}}_{\mathcal{T}}, \underbrace{\overline{L}^{(s)}, \ldots, \overline{L}^{(s)}}_{\mathcal{T}}\right]^{\top}, \quad (4)$$

where $\overline{T}^{(s)}$ and $\overline{L}^{(s)}$ are respectively the upper bound of the setting value of IDT for the AC and AMI for the LT.

$$D_{i,t} = f_D\left(D_{i,t-1}, D_{i,t-1}^+, D_{i,t-1}^-, T_t, T_t^+, T_t^-, L_t, L_t^+, L_t^-, E_{i,t}\right), \quad (5)$$

where $f_D$ is a prediction model for DL formulated as a linear regression model; $T_t$ and $L_t$ are respectively the time-averaged prediction value of IDT and AMI at time step $t$-th. $E_{i,t}$ is awakening effort of worker $i$ at time step $t$-th and it is defined as standard deviation (SD) of instant values of DL within the time step. Notation $X_t^+$ represents the increment between time steps $t$-th and $(t-1)$-th defined as $max\{X_t - X_{t-1}, 0\}$. Similarly, notation $X_t^-$ represents the decrement between time step $t$-th and $(t-1)$-th defined as $max\{X_{t-1} - X_t, 0\}$. It is defined here that the start time for executing the calculation of D-MPC is within time step $t = 0$. Instead of predicted values, estimated and/or measured values are used for $D_{i,t}$, $T_t$, $L_t$, and $E_{i,t}$ when $t \leq 0$. In addition, it is assumed that $E_{i,t} = E_{i,0}, \forall t \in \{1,2,\ldots,\mathcal{T}\}$. Note that explanatory variables $T_t^+$, $T_t^-$, $L_t^+$, and $L_t^-$ play an important role for the control behavior of D-MPC changing IDT/AMI from low to high and vice versa.

$$T_t = f_T(T_{t-1}, T_t^{(s)})$$
$$= \begin{cases} k^+ T_t^{(s)} + (1-k^+)T_{t-1} & T_t^{(s)} \geq T_{t-1} \\ k^- T_t^{(s)} + (1-k^-)T_{t-1} & otherwise \end{cases} \quad (6)$$

where $f_T$ is a prediction model of IDT formulated as an approximate model of a first-order lag system; $k^+$ and $k^-$ are coefficients regarding raising and lowering of IDT, respectively. This prediction model also plays an important role because IDT change rate which relates degree of stimulus can be accurately considered.

$$L_t = f_L(L_{t-1}, L_t^{(s)}), \quad (7)$$

where $f_L$ is a prediction model of AMI formulated as a linear regression model.

$$p_T |T_t - T^{(c)}| + p_L |L_t - L^{(c)}| \leq P, \forall t \in \{1,2,\ldots,\mathcal{T}\}, \quad (8)$$

where $T^{(c)}$, $L^{(c)}$, $p_T$, $p_T$, and $P$ are configured in advance as setting parameters. Here, $T^{(c)}$ and $L^{(c)}$ are respectively comfortable IDT and AMI; $p_T$ and $p_L$ are respectively comfort-penalty coefficients regarding degree of comfort violation defined as absolute error between actual and comfort values, namely $|T_t - T^{(c)}|$ and $|L_t - L^{(c)}|$. Hence, (8) represents the constraint where the total of normalized individual comfort penalties of different comfort factors (i.e., IDT and AMI) should be less than or equal to the configured value $P$. Therefore, this constraint is expected to be a function preventing IDT and AMI from becoming "uncomfortable" values at the same time.

The mathematical programming problem defined as (1)-(8) can be solved by differential evolution [15][16] which is one of optimization algorithms, and $\boldsymbol{U}^*$ is obtained as a result of optimization. The optimization is executed and the obtained $\boldsymbol{U}^*$ is sent to the AC and the LT as the setting values at the same constant interval $\tau$ as the size of the time step.

III. EVALUATION METHOD

A. Protocol

To evaluate the effectiveness of D-MPC, an experiment were conducted with healthy subjects, with uniform age and sex, who were selected and dispatched by a different company from that of the authors. Note that the authors and the subjects are not in a relationship creating a conflict of interest. The author explained to the subjects the contents of the experiment where their biological information and performance at work were measured under varied conditions of IDT and AMI. However, the subjects were not given the information about when IDT and AMI were changed and how their performance could change during the experiment. Besides, their activities during off-time were not limited.

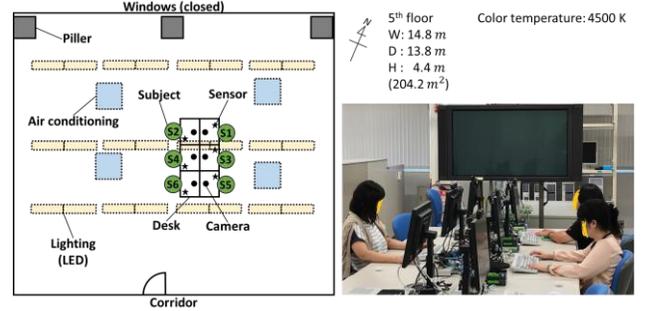

Figure 3. Experiment Environment

TABLE I. DETAILED INFORMATION OF SUBJECT.

| Subject ID | | S1 | S2 | S3 | S4 | S5 | S6 |
|---|---|---|---|---|---|---|---|
| Sex | | F | M | M | F | F | M |
| Age | [y/o] | 37 | 30 | 42 | 48 | 41 | 23 |
| HT | [cm] | 150 | 165 | 175 | 160 | 150 | 175 |
| WT | [kg] | 50 | 65 | 65 | 60 | 50 | 55 |
| Exp | [month] | 1.2 | 0 | 0 | 3.2 | 1.2 | 0 |

TABLE II. SCHEDULE OF EXPRIMENT.

| Experiment | Control | Date (in 2018) | Days |
|---|---|---|---|
| Case 1 | NOC | Sep: 4th,18th~21st | 5 |
| | MPC-1 | Sep: 6th,25th~28th | 5 |
| | MPC-2 | Sep: 7th,11th~14th | 5 |
| Case 2 | NOC | Dec: 3rd~7th,19th,20th | 7 |
| | MPC-1 | - | 0 |
| | MPC-2 | Dec: 10th~14th,17th,18th | 7 |

Table I shows detailed information about the six subjects. Subjects S1, S4, and S5 had experience of the task described in the next subsection. They were proficient at the task because an average person can master the task within four to five days.

Figure 3 shows the experiment environment. It is in the facility of the author, but we did not enter it during the experiment to reduce psychological influence of us on the subjects.

Table II shows the schedule of the experiment. Case 1 targets a cooling operation in the summer, and Case 2 targets a heating operation in the winter. In Table II, NOC represents the case of no control of AC/LT, MPC-1 represents the case of D-MPC applied to AC only, and MPC-2 represents the case of D-MPC applied to both AC and LT. The setting values of IDT for AC and brightness for LT were respectively kept constant as $T^{(c)}$ and $L^{(c)}$ in the case the AC/LT were not controlled. To reduce the influence of proficiency of the task, acclimation to the experiment environment, and psychological effects (e.g., primacy/recency effect), the subjects additionally did the task for five workdays before the experiment period and for over three workdays after it. IDT and AMI were changed day-by-day for five days before the experiment period. The purpose of this procedure is prior experience to reduce psychological effects due to being unacclimated to change of IDT and AMI.

B. Task and work condition

As a real routine work, the subjects did the labeling task[5] mentioned in a previous study [6]. The sequence of the labeling task is as follows: first, viewing a facial video for 10 seconds; second, evaluating DL of the facial video on a scale of 1 to 5 in accordance with the criterion shown in Table III; and third, inputting the evaluated value into a spreadsheet. Before the experiment, substantial amount of facial video was prepared, and individual facial video was shuffled so as not to be evaluated in series of the same face and same time period.

---

[5] Labeling task is a kind of annotation that is a major and important task in research and development of technologies based on machine learning. It is not easily replaced by automation technologies and will exist for the foreseeable future.

The working time period was 10:00-17:00, but the subjects did not work in 12:00-13:00 as their lunch break. The subjects were allowed to take short breaks (e.g., for the toilet), absent themselves, and leave early for their own reasons. If a subject was late for any reason (e.g., delay of transportation), he/she immediately started working after arriving at the experiment environment.

In addition to the labeling task, the subjects answered a questionnaire via their web browser every hour on the hour. The questionnaire was regarding subjective evaluation on drowsiness, comfort, and so on.

*C. Evaluation index*

Evaluation indices regarding productivity, DL, and comfort are defined as follows.

In the evaluation, productivity is defined as averaged processing speed [labels/h] for each subject. Averaged processing speed was calculated by averaging processing speed for each workday. Processing speed was calculated as the value given by number of labels divided by actual working time in a day. Number of labels is the amount that a subject processed in the day, and actual working time is the time between the start and end of work except lunch break of one hour.

Two types of DL were evaluated. The first one is an estimated value calculated by the method used in a previous study [6] (learning data: 45 subjects; estimation accuracy: R=0.82/MAE=0.40). The second one is a subjective evaluation collected from the answers to the questionnaire. The subjects answered their subjective evaluation of their own DL in accordance with the criterion shown in Table III.

Two types of comfort were evaluated. Thermal comfort and comfort of brightness were subjectively evaluated in a similar way to that for evaluating the subjective DL. The subjects answered their subjective evaluation of their own comfort in accordance with the criterion shown in Table IV.

To reduce the influence of lack of sleep on accurate evaluation, the data obtained when the sleep time of a subject is less than or equal to four hours were excluded, because mean sleep time of the subjects is about six hours. Only data from S3 (Case 1: 3 days; Case 2: 4 days) were excluded.

*D. Parameters*

Table V and VI show the parameters of D-MPC configured for the experiment. Parameters $T^{(c)}$, $L^{(c)}$, $p_T$, $p_L$, $P$, $\underline{T}^{(s)}$, $\overline{T}^{(s)}$, $\underline{L}^{(s)}$, and $\overline{L}^{(s)}$ relating to comfort were determined to make the subjects comfortable in reference to international standards [17][18] and the results of the questionnaire. Parameters $\mathcal{T}$ and $\tau$ relating to time were determined by doing trial-and-error until reasonable control behavior was obtained.

TABLE III. CRITERION FOR DROWSINESS LEVEL.

| Drowsiness level | | Descriptions |
|---|---|---|
| 1: | Not drowsy at all (awake) | Fast and frequent gaze motions; stable eye blink; active body motions. |
| 2: | Slightly drowsy | Slow gaze motions; lips opening. |
| 3: | Drowsy | Slow and frequent eye blinks; posture adjusting; mouth moving; face touching. |
| 4: | Significantly drowsy | Frequent yawn; unnecessary body movements; slow eye blink or gaze motions; deep breaths. |
| 5: | Extremely drowsy | Eyelids closing; head tilting forward/backward. |

TABLE IV. CRITERION FOR SUBJECTIVE COMFORT LEVEL.

| Comfort level | Descriptions |
|---|---|
| 0 | Unaware; comfortable. |
| +1/-1 | Unaware when concentrating; Slightly hot/cold or bright/dark. |
| +2/-2 | Aware but tolerable; Moderately hot/cold or bright/dark. |
| +3/-3 | Intolerable; Significantly hot/cold or bright/dark. |

TABLE V. SETTINGS OF COMMON PARAMETERS.

| $\tau$ [h] | $\mathcal{T}$ | $T^{(c)}$ [°C] | $p_T$ [/°C] | $\underline{L}^{(s)}$ [lx] | $\overline{L}^{(s)}$ [lx] | $L^{(c)}$ [lx] | $p_L$ [/lx] | $P$ |
|---|---|---|---|---|---|---|---|---|
| 0.25 | 4 | 26.0 | 0.5 | 450 | 750 | 600 | 150 | 2 |

TABLE VI. SETTINGS OF PARAMETERS IN EACH CASE

| Experiment | $\mathcal{W}$ | $\underline{T}^{(s)}$ [°C] | $\overline{T}^{(s)}$ [°C] |
|---|---|---|---|
| Case 1 | 5 (S1~S5) | 25.5 | 26.5 |
| Case 2 | 6 (S1~S6) | 25.0 | 27.0 |

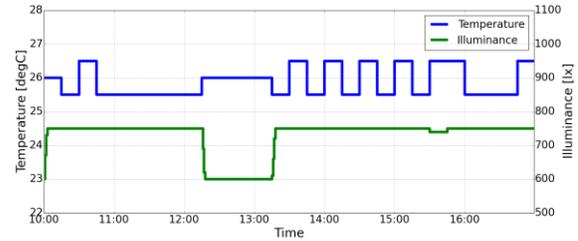

Figure 4. Result of setting values in Case 1 (11th Sep.).

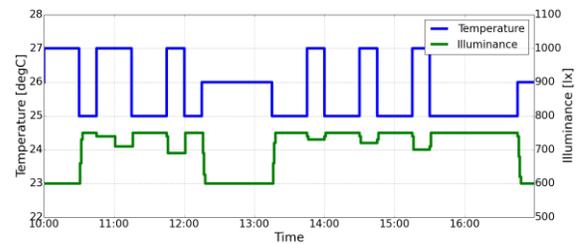

Figure 5. Result of setting values in Case 2 (14th Dec.).

The regression coefficients of $f_D$ are identified by a method of parameter estimation with learning data of 57 subjects in Case 1 and 51 subjects in Case 2. The model coefficients of $f_T$ (i.e., $k^+$ and $k^-$) were identified by a method of parameter estimation with learning data for the latest two weeks. The regression coefficients of $f_L$ are identified by a method of parameter estimation with learning data for one day when the setting value of LT was varied to cover the entire range of it.

## IV. EVALUATION RESULTS

### A. Results

Figure 4 and 5 respectively show examples of setting values determined by D-MPC (MPC-2) in Case 1 and Case 2. It can be seen that D-MPC dynamically determined setting values for AC/LT in both cases. The setting value for AC was not purely lowered but adaptively changed to minimize DL. In the same manner, the setting value of LT was not purely raised but also adaptively changed so as not to violate the comfort constraint. In Case 1, the setting value for AC was kept low so as to quickly lower the high IDT in the period 10:45-12:00, and so on when IDT was not lowered as usual because the AC was not working well for some reason (e.g., high outside temperature). In Case 2, the setting value for LT was occasionally lowered to satisfy the constraint condition (8) for maintaining comfort in the period 10:00-10:30, 15:15, and so on when IDT was too low because of the AC was not working well.

Table VII shows comparison of processing speeds of each subject in Case 1. In the cases of both MPC-1 and MPC-2, average processing speed of all subjects (appearing at row "ALL") tended to be improved. MPC-2 surpassed MPC-1 regarding speed-up, and that result means the combination of AC and LT could be more effective than AC alone. However, that is not always true for individual subjects. Subject S4 achieved higher processing speed in both MPC-1 and MPC-2; however, subject S2 and S5 achieved higher processing speed in MPC-1 than MPC-2. Similarly, Table VIII shows comparison of processing speeds of each subject in Case 2. It can also be seen that a slightly low but comparable level of improvement regarding processing speed was obtained by D-MPC.

Table IX shows comparison of DLs in Case 1. DL(est) and DL(sbj) in the table respectively mean estimated value of DL and subjective evaluation of DL. Subjective evaluation of DL was improved in the cases of MPC-1 and MPC-2; however, estimated value of DL was improved less. Similarly, Table X shows comparison of DLs in Case 2. Subjective evaluation of DL was improved less, and estimated valued of DL was slightly degraded. These results are discussed in the next subsection.

Table XI and XII respectively shows comparison of subjective comfort levels in Case 1 and Case 2. TH(sbj) and BR(sbj) in the table respectively mean rate of comfort response, that is defined as the rate between number of total answers and number of answers within the range from -1 to +1, regarding thermal comfort and the same of brightness comfort. NOC, MPC-1, and MPC-2 show almost no difference. In other words, D-MPC did not sacrifice comfort.

### B. Discussion

The causes of the improvement of DL as compared to processing speed are described hereafter.

First, it is presumed that control of AC/LT without sacrifice of comfort had small room for improvement of DL and the degree of the improvement was as much as the result of the experiment

Second, it was possible that even a small improvement of DL (<0.2) can considerably influence productivity.

TABLE VII. COMPARISON OF PROCESSING SPEED IN CASE 1

| Subject | NOC | MPC-1 | | MPC-2 | |
|---|---|---|---|---|---|
| | Mean (SD/N) | Mean (SD/N) | Speed-up [%] | Mean (SD/N) | Speed-up [%] |
| S1 | 534.0 (53.1/5) | 559.6 (66.9/2) | **+4.8** (p=.385) | 596.9 (64.2/4) | **+11.8** (p=.109) |
| S2 | 426.4 (69.9/4) | 469.7 (39.7/5) | **+10.2** (p=.195) | 423.4 (54.4/5) | **-0.7** (p=.477) |
| S3 | 355.8 (46.6/4) | 376.5 (26.2/4) | **+5.8** (p=.268) | 398.0 (16.8/4) | **+11.8** (p=.110) |
| S4 | 432.3 (13.4/4) | 482.3 (21.9/4) | **+11.6** (p=.010) | 512.0 (59.3/5) | **+18.4** (p=.027) |
| S5 | 394.7 (44.0/4) | 405.5 (41.6/3) | **+2.7** (p=.397) | 390.9 (39.8/5) | **-1.0** (p=.456) |
| ALL | 428.6 | 458.7 | **+6.8** (p=.007) | 464.2 | **+8.3** (p=.052) |

TABLE VIII. COMPARISON OF PROCESSING SPEED IN CASE 2

| Subject | NOC | MPC-1 | | MPC-2 | |
|---|---|---|---|---|---|
| | Mean (SD/N) | Mean (SD/N) | Speed-up [%] | Mean (SD/N) | Speed-up [%] |
| S1 | 427.0 (49.4/6) | - | - | 532.9 (185.4/7) | **+24.8** (p=.221) |
| S2 | 488.5 (18.5/7) | - | - | 502.6 (32.2/7) | **+2.9** (p=.188) |
| S3 | 386.3 (47.9/7) | - | - | 468.4 (39.5/3) | **+21.3** (p=.036) |
| S4 | 375.1 (20.0/6) | - | - | 423.0 (26.5/7) | **+12.8** (p=.003) |
| S5 | 385.4 (61.4/6) | - | - | 365.3 (69.6/6) | **-5.2** (p=.313) |
| S6 | 648.9 (62.9/6) | - | - | 594.6 (70.8/7) | **-2.6** (p=.348) |
| ALL | 445.4 | - | - | 481.1 | **+8.0** (p=.076) |

TABLE IX. COMPARISON OF DROWSINESS LEVEL IN CASE 1

| Index | NOC | MPC-1 | | MPC-2 | |
|---|---|---|---|---|---|
| | Mean (SD) | Mean (SD) | Difference | Mean (SD) | Difference |
| DL (est) | 1.72 (0.38) | 1.71 (0.39) | **+0.01** (p=.901) | 1.70 (0.37) | **+0.02** (p=.550) |
| DL (sbj) | 1.55 (0.84) | 1.35 (0.62) | **+0.20** (p=.015) | 1.45 (0.74) | **+0.10** (p=.238) |

TABLE X. COMPARISON OF DROWSINESS LEVEL IN CASE-2

| Index | NOC | MPC-1 | | MPC-2 | |
|---|---|---|---|---|---|
| | Mean (SD) | Mean (SD) | Difference | Mean (SD) | Difference |
| DL (est) | 1.53 (0.31) | - | - | 1.59 (0.34) | **-0.06** (p<.001) |
| DL (sbj) | 1.57 (0.89) | - | - | 1.54 (0.82) | **+0.03** (p=.698) |

TABLE XI. COMPARISON OF COMFORT LEVEL IN CASE 1

| Index | NOC | MPC-1 | | MPC-2 | |
|---|---|---|---|---|---|
| | Mean (SD) | Mean (SD) | Difference | Mean (SD) | Difference |
| TH (sbj rate) | 81.5% (38.8%) | 91.4% (28.0%) | **+9.9** (p=.008) | 79.7% (40.3%) | **-1.8** (p=.659) |
| BR (sbj rate) | 98.8% (10.8%) | 100.0% (0.0%) | **+1.2** (p=.158) | 100.0% (0.0%) | **+1.2** (p=.158) |

TABLE XII. COMPARISON OF COMFORT LEVEL IN CASE 2

| Index | NOC | MPC-1 | | MPC-2 | |
|---|---|---|---|---|---|
| | Mean (SD) | Mean (SD) | Difference | Mean (SD) | Difference |
| TH (sbj rate) | 90.1% (29.9%) | - | - | 92.1% (27.0%) | **+2.0** (p=.401) |
| BR (sbj rate) | 99.7% (5.7%) | - | - | 98.6% (11.9%) | **-1.1** (p=.160) |

Third, it was possible that accuracy and/or resolution of measured DL, especially in the range of 1 to 2, was not enough for the evaluation with the experiment. As the reason for that insufficiency of accuracy and/or resolution, it is presumed that it was difficult to estimate DL minutely from eyelid motion and to quantify DL precisely from recognition of the subject's own bodily state. However, it is presumed that D-MPC, especially the prediction model for DL, is not influenced much by the above-mentioned issue regarding accuracy and/or resolution of DL estimation. Specifically, it is presumed that there were less influence to identification of the regression coefficients of $f_D$ because of learning data. For the learning data of $f_D$, DL had distributed within in the range of 1 to 3, and the range wider than estimation error of DL (MAE=0.40) resulted in valid identification of the regression coefficients.

Fourth, it was possible that DL estimation accuracy was degraded in MPC-2 of Case 2 because relative humidity was 10-15% lower than the same in NOC due to outside condition. Lower relative humidity could increase eye blinking, resulting in overestimation of DL relative to actual DL.

For more reliable and consolidated evaluation on the relation between productivity and drowsiness, it is expected that the above-mentioned points are considered in future works.

## V. Conclusion

This paper proposed a methodology of model predictive control for alleviating shallow drowsiness of office workers and thus improving their productivity. The methodology is based on dynamically scheduling setting values for air conditioning and lighting to minimize drowsiness level of office workers on the basis of a prediction model that represents the relation between future drowsiness level and combination of indoor temperature and ambient illuminance. The prediction model was be identified by utilizing state-of-the-art drowsiness estimation method. The proposed methodology was evaluated in regard to a real routine task (performed by six subjects over five workdays), and the evaluation results demonstrated that the proposed methodology improved the processing speed of the task by 8.3% without degrading comfort of the workers.

## Ethical Consideration

This research was carried out under the approval of ethics review committee of the Japan Conference of Clinical Research (JCCR) specified non-profit organization with the best respect regarding ethical guidelines of clinical study, Declaration of Helsinki, other related laws/regulations, and individual information/health of subjects. All the experiments in the research were conducted after obtaining informed consents from the subjects

## References

[1] International Monetary Fund, "Technology and the Future of Work," 2018.
[2] O. Seppanen, W. J. Fisk, and D. Faulkner, "Control of Temperature for Health and Productivity in Offices," *ASHARE Transactions*, vol. 111, pp.680-686, 2005.
[3] F. Obayashi, M. Kawauchi, M. Terano, K. Tomita, Y. Hattori, H. Shimoda, H. Ishii, and H. Yoshikawa, "Development of an Illumination Control Method to Improve Office Productivity," *Int'l Conf. on Human-Computer Interaction (HCI)*, pp.939-947, July 2007.
[4] T. Shinozuka, T. Ikaga, C. Kaseda, M. Miura, and K. Mizutani, "Office Occupant Productivity under Variable HVAC Control Based Thermal Satisfaction," *Int'l Conf. on Healthy Buildings*, pp. 1611-1616, July 2012.
[5] T. Warita, T. Ikaga, K. Harimoto, and M. Ichihara, "Effect of Illuminance and Color Temperature on Productivity," *Int'l Conf. on Indoor Air Quality and Climate*, pp. 1785-1790, June 2011.
[6] M. Tsujikawa, Y. Onishi, Y. Kiuchi, T. Ogatsu, A. Nishino, and S. Hashimoto, "Drowsiness Estimation from Low-Frame-Rate Facial Videos Using Eyelid Variability Features," *Ann. Int'l Conf. of the IEEE Engineering in Medicine and Biology Society (EMBC)*, pp. 5203-5206, July 2018.
[7] M. Sun, M. Tsujikawa, Y. Onishi, X. Ma, A. Nishino, and S. Hashimoto, "A Neural-Network-Based Investigation of Eye-Related Movement for Accurate Drowsiness Estimation," *Ann. Int'l Conf. of the IEEE Engineering in Medicine and Biology Society (EMBC)*, pp.5207-5210, July 2018.
[8] V. E. Wilkinson, M. L. Jackson, J. Westlake, B. Stevens, M. Barnes, S. M. W. Rajaratnam, and M. E. Howard, "The Accuracy of Eyelid Movement Parameters for Drowsiness Detection," *Journal of Clinical Sleep Medicine (JCSM)*, vol. 9, no. 12, pp. 1315-1324, Dec. 2013.
[9] M. I. Chacon-Murguia and C. Priet-Resendiz, "Detecting Driver Drowsiness: A survey of system designs and technology," *IEEE Consumer Electronics Magazine*, vol. 4, pp. 107-119, Oct. 2015.
[10] J. Gwak, M. Shino, and M. Kamata, "Interaction between Thermal Comfort and Arousal Level of Drivers in Relation to the Changes in Indoor Temperature," *Int'l Journal of Automotive Engineering*, vol.9, no. 2, pp.86-91, June 2018.
[11] S. Benedetto, A. Carbone, V. Drai-Zerbib, M. Pedrotti, and T. Baccino, "Effects of luminance and illuminance on visual fatigues and arousal during digital reading," *Computers in Human Behavior*, vol. 41, pp.112-119, Dec. 2014.
[12] E. Aidman, C. Chadunow, K. Johnson, and J. Reece, "Real-time driver drowsiness feedback improves driver alertness and self-reported driving performance," *Accident Analysis and Prevention*, vol. 81, pp. 8-13, Apr. 2015.
[13] D. Q. Mayne, J. B. Rawlings, C. V. Rao, and P. O. M. Scokaert, "Constrained model predictive control: Stability and optimality," *Automatica*, vol. 36, pp. 789-814, June 2000.
[14] J. Cigler, S. Privara, Z. Vana, D. Komakova, and M. Sebek, "Optimization of Predicted Mean Vote Thermal Comfort Index within Model Predictive Control Framework," *IEEE Conf. on Decision and Control (CDC)*, pp. 3056–3061, 2012.
[15] R. Storn and K. Price, "Differential Evolution – A Simple and Efficient Heuristic for Global Optimization over Continuous Spaces," *Journal of Global Optimization*, vol. 11. pp. 341–359, 1997.
[16] S. Das and P.N. Suganthan, "Differential Evolution: A Survey of the State-of-the-Art," *IEEE Trans. on Evolutionary Computation*, vol. 14, no. 1, pp. 4–31, 2011.
[17] ISO7730:2005, Ergonomics of the thermal environment -analytical determination and interpretation of thermal comfort using calculation of PMV and PPD indices and local thermal comfort criteria, ISO, 2005.
[18] ISO8995:2002, Lighting of Indoor Work Places, ISO, 2002.
[19] S. Ohly, S. Sonnentag, and F. Pluntke, "Routinization, work characteristics and their relationships with creative and proactive behaviors," *Journal of Organizational Behavior*, pp.257-279, 2006
[20] A. Bruggen, C. Feichter, and M. G. Williamson, "The effect of input and output targets for routine tasks on creative task performance," *The Accounting Review*, vol. 93, pp.29-43, Jan. 2018.